\documentclass[english,aps,prd,nofootinbib,showkeys,preprint,floatfix,notitlepage]{revtex4}
\pdfoutput=1
\usepackage[utf8]{inputenc}
\usepackage[english]{babel}
\usepackage{amsmath}
\usepackage{amssymb}
\usepackage{mathtools}
\usepackage{graphicx}
\usepackage{multirow} 
\usepackage[colorlinks=True]{hyperref}
\usepackage{lmodern}
\usepackage{wasysym}

\def\gsim{\gtrsim}
\def\lsim{\lesssim}

\newcommand{\AddrUsM}{
{\it Universidad T\'ecnica Federico Santa Mar\'ia and Centro Cient\'ifico Tecnol\'ogico de Valparaiso,\\
    Casilla 110-V, Valparaiso, Chile.}}

\newcommand{\AddrUdeA}{%
 Instituto de F\'\i sica, Universidad de Antioquia,\\
 Calle 70 No. 52-21, Medell\'in, Colombia. }

\newcommand{\AddrAHEP}{
  {\it AHEP Group, Instituto de F\'{\i}sica Corpuscular --
    C.S.I.C./Universitat de Val{\`e}ncia \\
    Edificio de Institutos de Paterna, Apartado 22085,
  E--46071 Val{\`e}ncia, Spain}}

\begin{document}

\preprint{IFIC/17-16}  

\title{Fermionic triplet dark matter in an $SO(10)$-inspired left right model}

\author{Carolina Arbel\'aez R.} \email{carolina.arbelaez@.usm.cl} \affiliation{\AddrUsM}
\author{Martin Hirsch}\email{mahirsch@ific.uv.es} \affiliation{\AddrAHEP}
 \author{Diego Restrepo}\email{restrepo@udea.edu.co} \affiliation{\AddrUdeA}









\begin{abstract}
We study a left right (LR) extension of the Standard Model (SM) where the Dark Matter (DM) candidate is composed of a set of fermionic Majorana triplets. The DM is stabilized by a remnant $Z_{2}$ symmetry from the breaking of the LR group to the SM. Two simple scenarios where the DM particles plus a certain set of extra fields lead to gauge coupling unification with a low LR scale are explored. The constraints from relic density and predictions for direct detection are discussed for both scenarios. The first scenario with a $SU(2)_R$ vectorlike fermion triplet contains a DM candidate which is almost unconstrained by current direct detection experiments. The second scenario, with an additional $SU(2)_R$ triplet, opens up a scalar portal leading to direct detection constraints which are similar to collider limits for right gauge bosons. The DM parameter space consistent with phenomenological requirements can also lead to successful gauge coupling unification in a  $SO(10)$ setup. 
\end{abstract}

\maketitle

\section{Introduction}
\label{sec:intro}
 
 To guarantee the stability of the dark matter, many models postulate a discrete symmetry, usually a $Z_2$, under which the standard model particles are even, while the dark matter is odd.\footnote{Recently also $Z_3, ... Z_N$ symmetries as the origin of the DM stability have been discussed, see for example \cite{Belanger:2012zr,Belanger:2012vp}.}  
From a theoretical point of view much more attractive would be of course, if such a symmetry had a deeper origin or at least some other phenomenological consequences apart from stabilizing the DM. An example for the former is a broken gauge symmetry. An example for the latter are discrete family symmetries, in which the stability of the DM is related to the generation of neutrino masses \cite{Hirsch:2010ru,Boucenna:2011tj}. 

One example of a discrete symmetry, which can emerge from the spontaneous breaking of a gauge symmetry is matter parity, $P_M=(-1)^{3(B-L)}$. In $SO(10)$ based models this discrete symmetry can survive breaking of $SO(10)$ and stabilize the dark matter as has been shown in \cite{Kadastik:2009dj,Kadastik:2009cu,Frigerio:2009wf}.  $SO(10)$ can be broken to the standard model group directly or in different steps. Interestingly, one of the intermediate groups that can arise from $SO(10)$ is the left-right (LR) symmetric group \cite{Senjanovic:1975rk}, $SU(3)_c\times SU(2)_L \times SU(2)_R \times U(1)_{B-L}$.  In the minimal LR model, gauge coupling unification can be achieved, if the LR scale is in the range of $\Lambda \simeq 10^{(10-11)}$~GeV \cite{Brahmachari:1991np}. However, for such a large scale, no phenomenological effects of the LR symmetry can be seen in DM -- apart from the stabilization of the DM candidate itself. However, it is possible to build models in which the LR scale can be lowered to the electro-weak scale, without destroying gauge coupling unification \cite{Arbelaez:2013nga}.

Such low-scale LR models can maintain an unbroken $Z_2$ after symmetry breaking, if the field that breaks $SU(2)_R \times U(1)_{B-L}$ has even charge under ($B-L$) \cite{Martin:1992mq}. Different models of this kind have been studied recently.  For example, the singlet component of a scalar ${\bf 16}$ as DM candidate has been studied in \cite{Mambrini:2016dca}.  Dark matter phenomenology in low-scale left-right symmetric models with fermionic  triplets ($\Psi_{1130}$ and $\Psi_{1310}$) has been studied in~\cite{Heeck:2015qra} along with quintuplets (also studied in~\cite{Agarwalla:2016rmw}) and in\cite{Garcia-Cely:2015quu} where also bidoublets  (see also ~\cite{Boucenna:2015sdg}), and scalar doublets or septets are studied \footnote{A note on notation: We use the transformation properties/charges of the fields under the LR and SM group to identify the fields: $\Psi_{SU(3)_c, SU(2)_L, SU(2)_R, U(1)_{B-L}}$}.

Potentially realistic models containing the $SO(10)$ ${\bf 10}$ and ${\bf 45}$ fermionic representations, from which a neutralino-like mass matrix with arbitrary mixings can be obtained, was discussed in \cite{Arbelaez:2015ila}. A model with a fermionic  (right-handed) 5-plet was studied in \cite{Agarwalla:2016rmw}.
Other examples in this line of thought include asymmetric dark matter from $SO(10)$ \cite{Nagata:2016knk}, or right-handed neutrinos as DM in a  ``dark left-right model", stabilized by an extra symmetry $S$ \cite{Dey:2015bur}. One can also explore different intermediate gauge groups from SO(10) and their connection to DM models, as has been done in \cite{Mambrini:2015vna}. Another example with right-handed neutrinos as DM based on the group $SU(3)_C\times SU(2)_L \times SU(2)_R \times U(1)_{Y_L} \times U(1)_{Y_R}$ can be found in  \cite{Dev:2016xcp,Dev:2016qeb}.

Unlike many studies based on non-supersymmetric left right SM extensions where the DM is considered as a single, unmixed state, as for example, pure fermion triplet, bidoublet etc \cite{Heeck:2015qra,Garcia-Cely:2015quu}, in this work we study some models where the DM candidate is not necessarily a pure state, but may instead be a mixture of two or more multiplets, similar to neutralinos in supersymmetry. In \cite{Berlin:2016eem,Patra:2015qny,Arbelaez:2015ila}, some fermion mixed DM model composed by combinations such as singlet-triplet, singlet-bidoublet,  triplet-bidoublet were already studied. Here, we consider a combination of $SU(2)_R$ triplet-triplet fermionic DM multiplets. Mixing between the fermions is induced through the coupling to the triplet scalar $\Delta_{R}$.  A similar model was studied in the context of the diphoton excess in~\cite{Berlin:2016hqw} for low values of the $\Psi_{1132}\oplus {\bar\Psi}_{113-2}$  triplet mass. 

In this paper, we extend the generic left right model with a certain set of fields, see next section, such that gauge couplings unify in the ballpark of $m_{G}\simeq 2\times 10^{16}$~GeV. The dark matter in our setup are right-handed fermionic triplets. We start with a very simple scenario, denoted as \emph{Case I}, with a DM candidate from a vector-like pair of fields, $\Psi_{1132}$ and ${\bar\Psi}_{113-2}$. A second scenario, denoted as \emph{Case II}, contains a mixed DM candidate build from the above fields plus a $\Psi_{1130}$. The addition of the latter allows to add a scalar portal interaction to the model and a non-zero direct detection cross section $\sigma_{N}^{\text{SI}}$ appears. Note that \emph{Case I} arises from \emph{Case II} as a specific limit on the DM masses. We also check that both setups can lead to successful gauge coupling unification and calculate the parameter space allowed by proton decay constraints.

The rest of this paper is organized as follows. In Sec. \ref{sec:setup} we start by recalling the basics of the minimal left right model and $SO(10)$ inspired unification. To have successful unification of gauge couplings (GCU) and at the same time a ``low" LR scale (i.e. order TeV), requires additional fields. We discuss a particular set of fields  (``configuration") which gives correct GCU and also fulfill some additional phenomenological requirements. We identify the GUT parameter space and the region of DM candidate masses where GCU and successful fermionic DM simultaneously arise in Sec. \ref{sec:gut}. In Sec. \ref{sec:fermionicDM} we discuss the DM phenomenology of our two simple cases of fermionic DM in more detail. The relic density and direct detection cross sections are calculated. Finally in Sec. \ref{sec:conclusions} we conclude with a discussion of our results. Some technical aspects of our work are presented in the appendix.

\section{Model framework}
\label{sec:setup}

In this section we will briefly describe a non-SUSY left-right symmetric scenario inspired by $SO(10)$-like gauge coupling unification. As mentioned above, breaking the $U(1)_{B-L}$ symmetry by a field with even charge can leave a remnant $Z_2$ symmetry in the model, which allows to stabilize the DM. First, let us consider possible DM candidates. Considering multiplets up to \textbf{144} representations, see Appendix, we can find several scalar and fermionic candidates. First note that scalar candidates need to be odd with respect to $(B-L)$, because all standard scalars are even. Thus, scalar candidates can be found in the \textbf{16} or the \textbf{144}. Fermionic DM candidates, on the other hand, have to be $(B-L)$ even, since all SM fermions are odd under $(B-L$). Thus, possible candidates can be found in the \textbf{1}, \textbf{10}, \textbf{45}, $\cdots$ \textbf{126}. In our numerical study, we will concentrate on the DM candidates found in the \textbf{45} ($\Psi_{1130}$) and \textbf{126}/$\overline{\bf 126}$ ($\Psi_{1132}/{\bar\Psi_{113-2}}$) multiplets. We will study two cases, denoted as \emph{Case~I} and \emph{Case~II}. In \emph{Case~I}, the DM is taken to be the neutral component of the vector-like pair which belongs to the $SO(10)$ representations $\Psi_{1132}/{\bar\Psi_{113-2}}$. This scenario is very minimal in the sense that it has only one additional new parameter, which corresponds to the bare mass term of the vector-like pair and also gives a zero spin-independent (SI) DM-nucleon cross section at tree-level, as will be discussed in the next section. In \emph{Case~II}, we  add to the above fermion vector-like pair an extra $SU(2)_{R}$ fermionic triplet which belongs to the \textbf{45} representation, i.e. $\Psi_{1130}$. In this case, new terms, can be added to the Lagrangian, mixing the different neutral components in $\Psi_{1132}/{\bar\Psi_{113-2}}$ and $\Psi_{1130}$. A nonzero SI DM-nucleon cross section results. As expected, the $\emph{Case~I}$ is recovered in the limit where the DM particles are unmixed states and   $\Psi_{1130}$ is decoupled. 

Although our numerical calculations are done in a left-right symmetric model, the underlying theory at higher energies should be unifiable into $SO(10)$. We thus consider constraints arising from gauge coupling unification. For this, we consider a simple configuration of fields which contains our DM candidates, but adds a few more fields, such that the gauge couplings unify correctly at a $m_{G}$ scale allowed by proton decay. We then discuss the allowed GUT parameter space of this setup.

\subsection{Left-Right scalar sector}


The first stage of the symmetry breaking $SO(10) \to$ LR arises when a scalar field belonging to the \textbf{54} of $SO(10)$ representation acquires a vacuum expectation value (vev). Although our analysis is inspired by such a $SO(10)$ unification, we do not concern ourselves in detail with this first step. The second step is to break the LR group to the SM, which is then broken to $U(1)_{\text{EM}}$. In the minimal LR scenario, the scalar sector consist of only two multiplets: a bi-doublet $\Phi_{1220}$, needed to give correct masses to the electroweak vector bosons and SM charged fermions, and a scalar triplet $\Delta_{R} \equiv \Phi_{113-2}$, which breaks the LR group to the SM one. The neutral and charged components of these multiples can be written as \cite{PhysRevD.22.2227}: 
\begin{equation}
\Phi = 
 \begin{pmatrix}
  \Phi_{1}^{0} & \Phi_{2}^{+}  \\
  \Phi_{1}^{-} & \Phi_{2}^{0} \\
 \end{pmatrix}, \hspace{0.8cm}
\Delta_{R} = 
 \begin{pmatrix}
  \Delta_{R}^{-}/\sqrt{2} & \Delta_{R}^{--}  \\
  \Delta_{R}^{0} & -\Delta_{R}^{-}/\sqrt{2} \\
 \end{pmatrix}.
\end{equation}
It is assumed that the neutral components of these fields acquire vevs:
\begin{equation}
\langle\Phi\rangle=\begin{pmatrix}
 v_{1} & 0  \\
 0 & v_{2} e^{i\alpha} 
 \end{pmatrix}, \hspace{0.8cm}
 \langle\Delta_{R} \rangle = \begin{pmatrix}
  0 & 0  \\
 v_{R} & 0 
 \end{pmatrix}.
 \label{eq:vevs}
\end{equation}
The parameters $v_{1,2}$ are real and positive. For more details on left-right symmetry and gauge boson masses see \cite{Mohapatra:1986uf}. In our analysis, we do not assume an exact LR symmetry, i.e. $g_{L} \neq g_{R}$, see below.

\subsection{Gauge coupling unification constraints}
\label{sec:gut}

\noindent 
In this subsection, we will discuss briefly gauge coupling unification (GCU) and possible constraints on the parameter space of LR dark matter models. As is well-known \cite{Brahmachari:1991np}, the minimal LR model can lead to GCU only if the LR scale is 
of the order of $10^{10-11}$~GeV. In order to lower this scale to a phenomenologically interesting range, additional particles need to be added to the minimal model. We will use the results of \cite{Arbelaez:2013nga}. Essentially, we require the following two conditions to be fullfilled:
\begin{itemize}
\item[\textbf{(i)}] \emph{Perturbative unification:} This implies that the  gauge couplings unify with a value of $\alpha_{G}$ in the perturvative regime. Since our simple calculation does not consider GUT-scale thresholds, we are not necessarily imposing exact unification of the gauge couplings at the GUT scale ($m_{G}$). Rather, we allow for a difference of the gauge couplings at $m_{G}$ falling into a ``small nonunification triangle", i.e: $\alpha_{3}(m_{G})-\alpha_{2}(m_{G})\lesssim 0.9$ \cite{DeRomeri:2011ie, Arbelaez:2013hr}.
\item[\textbf{(ii)}] \emph{Proton decay:}
In non-supersymmetric $SO(10)$ GUTs models, the primary mode of proton decay is $p\rightarrow \pi^{0}e^{+}$. We consider the model valid if, in all the parameter space, it fulfills the constraint from proton decay $\tau_{p\rightarrow \pi^{0}e^{+}} \gtrsim 10^{34}$ years \cite{Nath:2006ut,Abe:2013lua}. The gauge $d=6$ operator associated to this decay leads a GUT scale of  $m_{G}^{4}\approx\tau_{p\rightarrow \pi^{0}e^{+}}\alpha_{G}^{2}m_{p}^{5}$. The current value of $\tau_{p\rightarrow \pi^{0}e^{+}}$ yrs. sets a lower limit on the GUT scale  of the order of $m_{G} \gtrsim 5\times 10^{15}$~GeV.

\end{itemize}

An extra set of fields added at an intermediate LR scale, denoted here as $M_{\text{LR}}\sim v_{R}$, gives new contributions to the $\beta$-coefficients of the gauge couplings. Many solutions that achieve GCU exist \cite{Arbelaez:2013nga}, but all of them require to add particles which transform non-trivially under color. For the numerical study we choose the following set of fields:
\begin{align}
\text{SM} &+\Phi_{1220}+\Phi_{113-2} \nonumber \\
   &+{\color{blue}\Psi_{1130}}+{\color{blue}\Psi_{1132}}+{\color{blue}{\bar\Psi}_{113-2}}\nonumber \\
   &+\Psi_{1310}+\Psi_{321\frac{1}{3}}+\bar{\Psi}_{321-\frac{1}{3}}+\Psi_{8110}\, .
\label{Config1}
\end{align}
The scalar bidoublet $\Phi_{1220}$ and the scalar triplet $\Phi_{113-2}$ are needed to achive the correct symmetry breaking pattern. The particles in the 2nd line are our dark matter candidates. In principle, also $\Psi_{1310}$ could be a dark matter candidate. Left right DM models with $\Psi_{1310}$ and $\Psi_{1130}$ as
possible DM candidates and $m_{\Psi_{1310}} \sim m_{\Psi_{1130}}$ have
already been studied in the literature \cite{Heeck:2015qra}. For the
case of $m_{\Psi_{1310}} \neq m_{\Psi_{1130}}$, the smaller of the two
will determine the character of the DM. If $\Psi_{1310}$ is the
lighter, results of \cite{Heeck:2015qra} will qualitatively apply
still.  We do not cover the mixed case with $\Psi_{1130}$
having a small component of $\Psi_{1310}$ in detail, because the
phenomenology will interpolate between these results.  Here
in this work, we show that a left triplet $\Psi_{1310}$ being heavier
than $\Psi_{1130}$ would not spoil GCU.

\begin{figure}[h]
\includegraphics[width=0.5\linewidth]{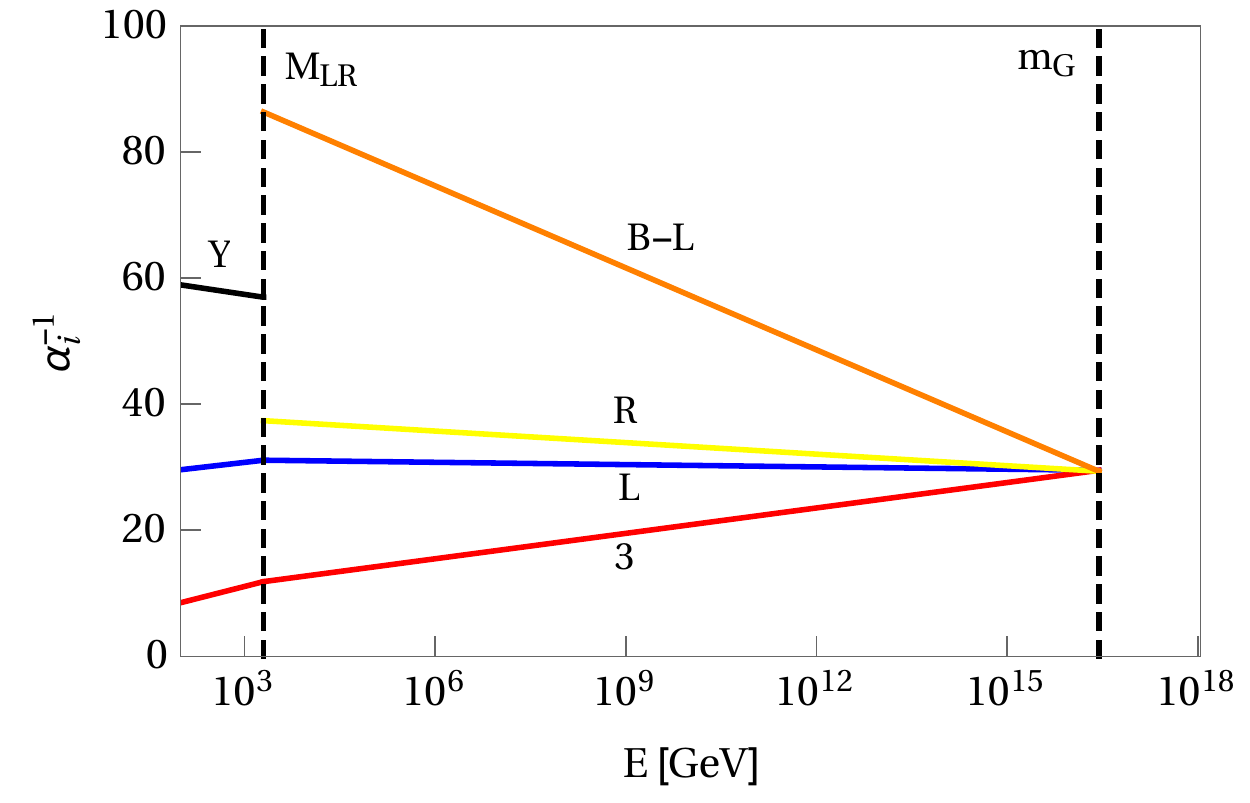}
\caption{Evolution of the gauge couplings for the configuration of fields described by Eq.~\eqref{Config1}, with $M_{\text{LR}}=2$~TeV. All new particle thresholds are added at $M_{\text{LR}}$ in this example.}
\label{fig:GCU1}
\end{figure}

The remaining colored fields are added to bring the prediction of $\alpha_S$ in agreement with experimental data. Note that all the extra fermionic fields can have vector-like masses. The evolution of the gauge couplings, explained in detail on Appendix B, corresponding to this configuration of fields is shown in Fig.~\ref{fig:GCU1}. There, all the new particle content, including the DM, is added at the scale $M_{\text{LR}}=2$~TeV.
Although ``exact parity'' ($g_{L} = g_{R} $) symmetry is
required in many constructions of LR models, this is not a mandatory
requirement for LR model building.  In particular, our model does not
have $g_{L} = g_{R}$ at the scale where the LR symmetry is
broken. Only for the sake of simplicity, we have chosen the number of
fields in our configuration as small as possible. Models with exact
parity (and a correspondingly larger set of fields) could easily be
constructed, without any fundamental changes in the phenomenology we
discuss here.

\begin{figure}[htb]
\begin{tabular}{cc}
\includegraphics[width=0.5\textwidth]{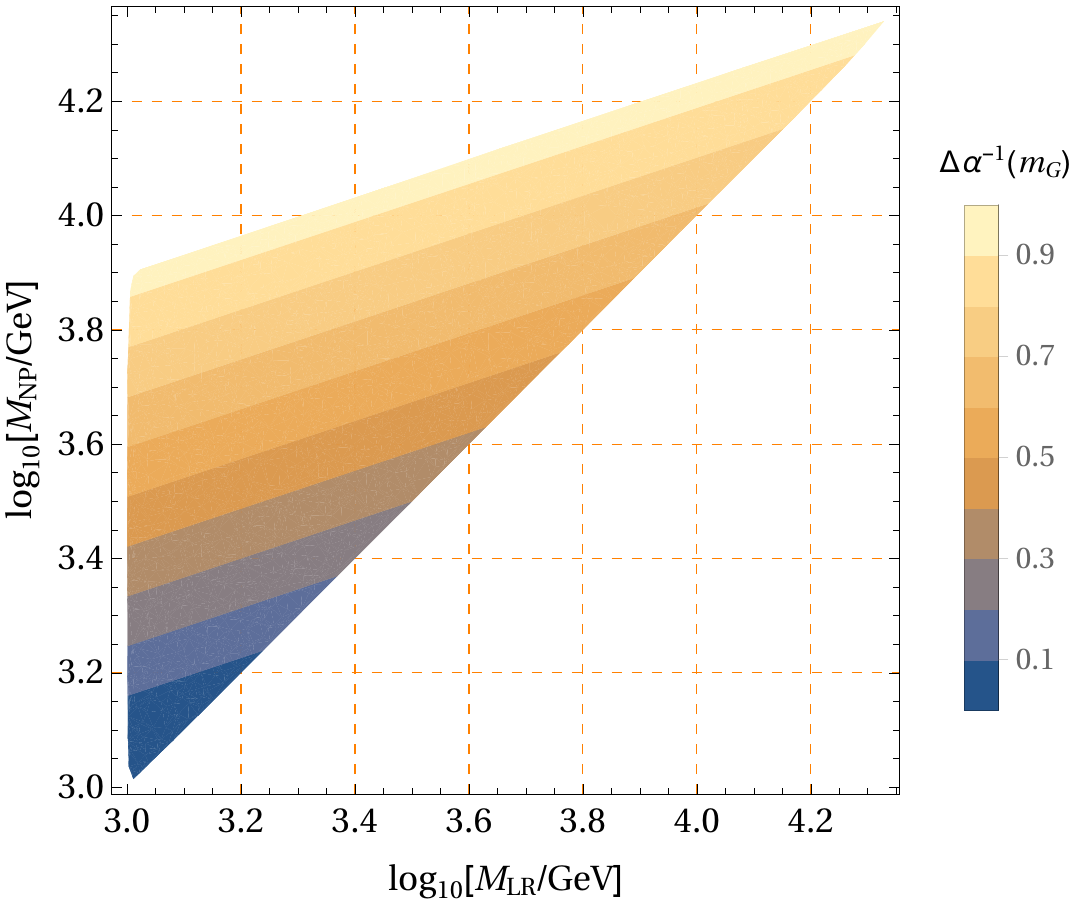} & %
\hspace{0.5cm}
\includegraphics[width=0.48\textwidth]{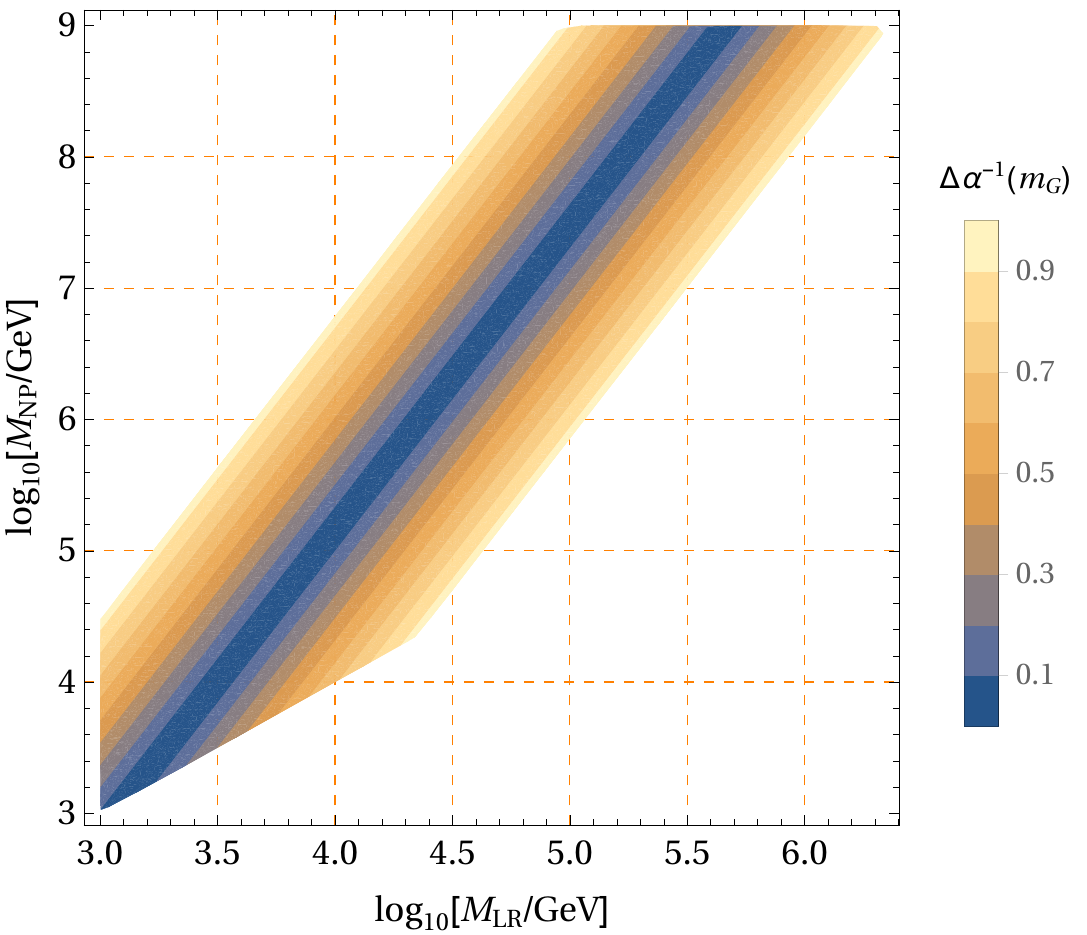}%
\end{tabular}
\caption{Allowed GUT parameter space passing the conditions $\textbf{(i)-(ii)}$. The scenarios  \textbf{(a)} (left) and \textbf{(b)} (right) are discussed in the text.}
\label{fig:GUTPS}
\end{figure}

An interesting question to ask is whether the requirement of correct GCU allows to constrain the mass scales of the model. Since the new fermions can all have vector-like masses, not necessarily related to the symmetry breaking scale $M_{\text{LR}}$, we will consider two simple scenarios: \textbf{(a)} Adding all the fields, including the fermion DM particles $\Psi_{1130}$, $\Psi_{1132}$ and ${\bar\Psi}_{113-2}$ at the scale $M_{\text{LR}}$, while the left triplet $\Psi_{1310}$ is added at some  scale of new physics denoted as $M_{\text{NP}}$. The resulting parameter space in the plane spanned by $M_{\text{LR}}$ and $M_{\text{NP}}$ is shown in the left panel of Fig.~\ref{fig:GUTPS}. And scenario 
\textbf{(b)}: only the scalar fields $\Phi_{1220}$ and $\Phi_{113-2}$ are added at the scale $M_{\text{LR}}$ while all the other fermions - including DM - are added at the $M_{\text{NP}}$ scale. The parameter space corresponding to this case is shown in the right panel of Fig.~\ref{fig:GUTPS}.

In both cases, the figures show contour lines for the size of the ``non-unification triangle", i.e. $\Delta(\alpha^{-1}(m_G))$ as a function of the new physics scales. As the figure to the left shows, in case \textbf{(a)} unification improves for low values of both $M_{\text{NP}}$ and $M_{\text{LR}}$ and values below $\Delta(\alpha^{-1}(m_G)) < 0.1$ requires the LR scale to be around 1~TeV. For this scenario, the LR symmetry breaking scale should be roughly below 20~TeV for $\Delta(\alpha^{-1}(m_G)) <0.9$. However, if we allow all new fermions to have masses larger than $M_{\text{LR}}$, case \textbf{(b)}, no upper limit on $M_{\text{LR}}$ can be inferred from this analysis, as the figure on the right shows. Note, however, that $M_{\text{NP}}$ has to be larger than $M_{\text{LR}}$ for good GCU to be maintained. Therefore for this case, constraints from the relic density provide interesting upper limits, as we will discuss in the next section. 

\section{Fermionic Dark Matter}
\label{sec:fermionicDM}


As a first step, we add to our LR minimal setup the two additional fermionic triplets $\Psi_{1132}\oplus {\bar\Psi}_{113-2}$  which represents a vector-like pair of Majorana DM. This scenario, which corresponds to a simple and unmixed DM case, is denoted  here as \emph{Case~I}. Considering that, in this scenario the vector-like  DM has zero hypercharge, the SI DM-nucleon cross section, $\sigma_{N}^{\text{SI}}$, is expected to be zero at tree level. As a second step, an extra  fermionic triplet $\Psi_{1130}$ is included to complete a scenario of mixed fermion Dark Matter in which, although the DM has hypercharge zero, a scalar portal interaction  of the DM through the interaction of the DM with the $\Delta_{R}$ generates a nonzero $\sigma_{N}^{\text{SI}}$.

All the numerical calculations of the next sections where implemented using SARAH \cite{Staub:2010jh,Staub:2008uz,Staub:2009bi} (based on the LR implementation in \cite{Staub:2016dxq}) which generates the necessary subroutines used subsequently by SPHENO \cite{Porod:2003um,Porod:2011nf}. The calculation of the relic density and the relevant cross sections is done by MicrOMEGAs~\cite{Belanger:2006is}, solving the Boltzman equation numerically through CalcHEP \cite{Belanger:2010st} output of SARAH. The scans were done using the SSP mathematica package \cite{Staub:2011dp}.

\subsection{Case I}

In this benchmark scenario  we introduced two Weyl fermions $\Psi_{1132}$ and ${\bar\Psi}_{113-2}$ which can be parametrized as:


%
\begin{equation}
\Psi_{1132} = 
 \begin{pmatrix}
  \Psi^{+}/\sqrt{2} & \Psi^{++}  \\
  \Psi^{0} & -\Psi^{+}/\sqrt{2} \\
 \end{pmatrix}, \hspace{0.8cm}
{\bar\Psi}_{113-2} = 
 \begin{pmatrix}
  \Psi^{-}/\sqrt{2} & \overline{\Psi}^{0}  \\
  \Psi^{--} & -\Psi^{-}/\sqrt{2} \\
 \end{pmatrix}.
\end{equation}

\begin{figure}[htb]
\begin{tabular}{ccc}
\includegraphics[width=0.3\textwidth]{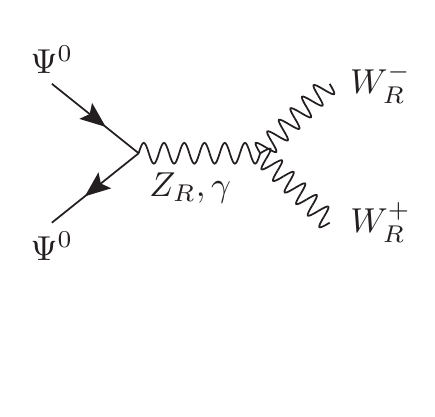} & %
\includegraphics[width=0.3\textwidth]{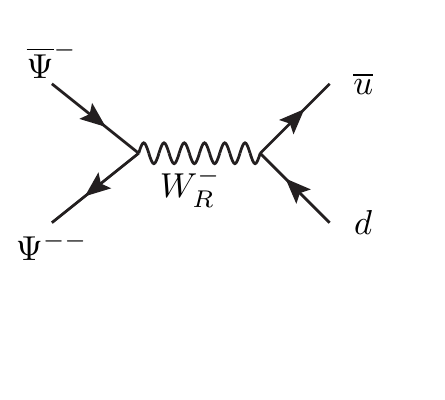} &
\includegraphics[width=0.3\textwidth]{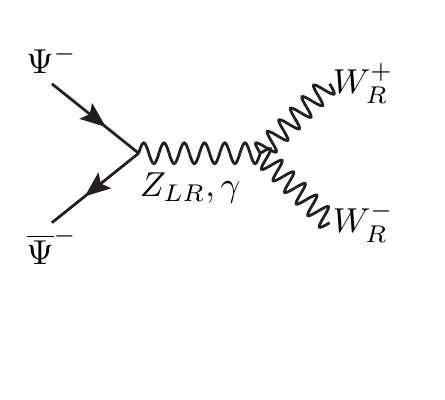}%
\end{tabular}
\begin{tabular}{ccc}
\includegraphics[width=0.3\textwidth]{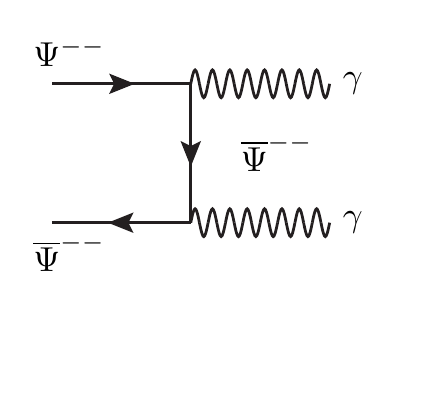} & %
\includegraphics[width=0.3\textwidth]{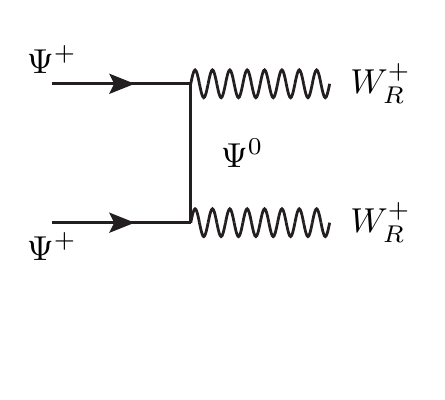}& 
\includegraphics[width=0.3\textwidth]{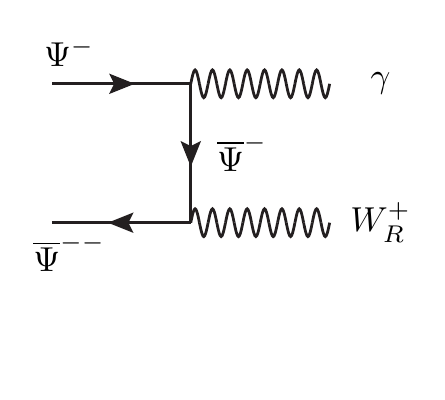}
\end{tabular}
\vspace{-1.5cm}
\includegraphics[width=0.3\textwidth]{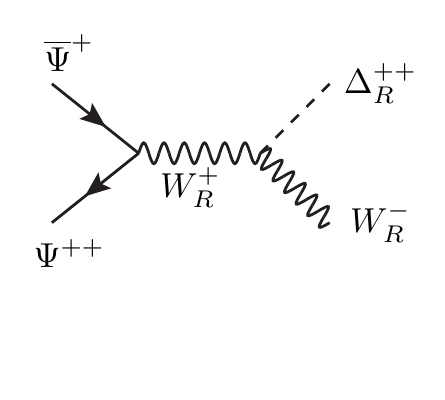}
\caption{Some of the Feynman diagrams for dark matter (co)annihilations determining the relic abundance of DM.}
\label{fig:Co}
\end{figure}

Note that, due to the quantum numbers and the chosen transformation properties of $\Psi_{1132}$ and ${\bar\Psi}_{113-2}$ under the LR gauge symmetry, the most general renormalizable Lagrangian contains only the following mass term as a new parameter: $L \supset M_{23}\operatorname{Tr}(\Psi_{1132}\bar\Psi_{113-2})$. $M_{23}$ corresponds to the tree level mass of the different $\Psi_{1132}$ and ${\bar \Psi}_{113-2}$ components. The absence of any interaction term mediating the decay of the DM particles into the SM particles guarantees that the lightest component of these triplets is accidentally stable and thus represents a DM candidate. In this scenario, the resulting relic density abundance $\Omega h^2$ depends not only on the DM mass, related directly by $M_{23}$, but also on the value of $v_{R}$, via the mass of $Z_{R}$ and $W_{R}$. By construction, in this setup the only interactions affecting the relic density abundance are the gauge interactions. When the mass splitting between the dark matter candidate and the charged components of the triplet are small, coannihilation effects need also to be included. This is done automatically in MicrOMEGAs. The most important annihilation and coannihilation processes contributing to the relic density are described in Fig.~\ref{fig:Co}. 


\begin{figure}[h]
\includegraphics[width=0.6\linewidth]{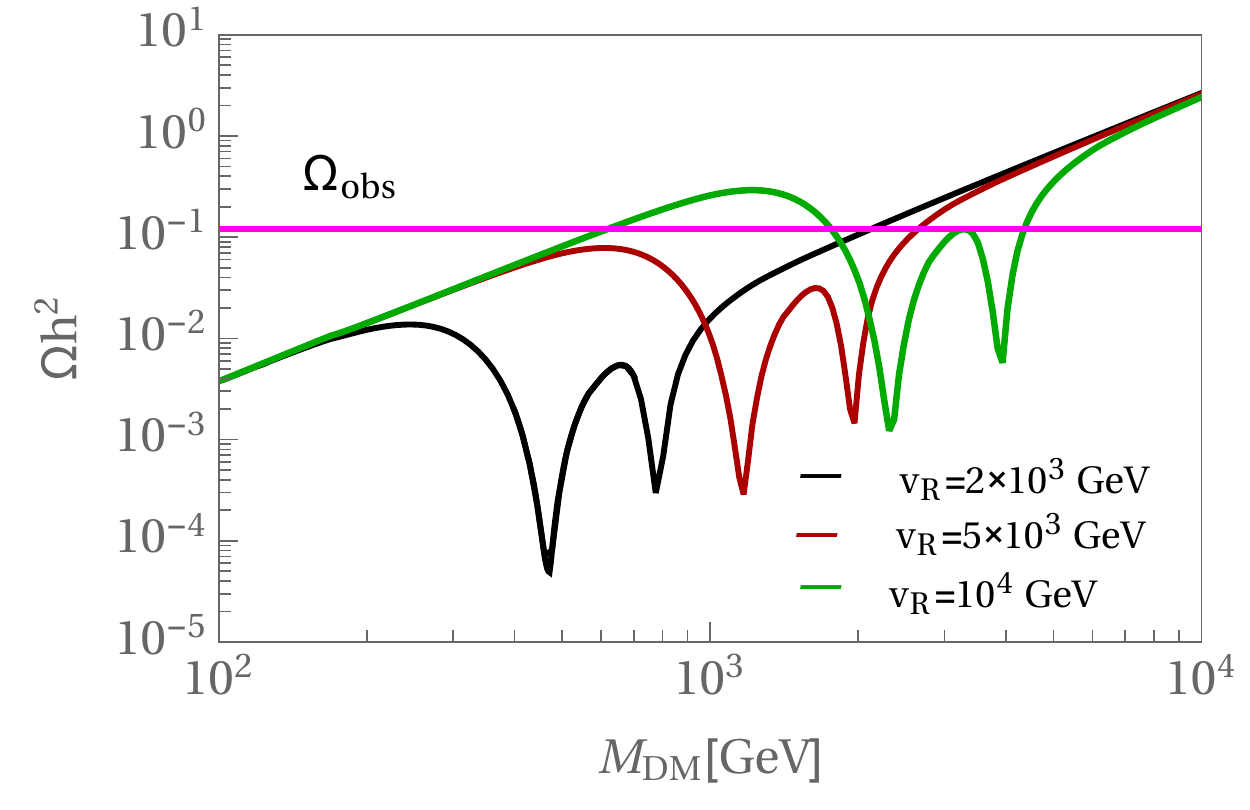}
\caption{Relic density $\Omega h^2$ as a function of $m_{\text{DM}}$ for different values of $v_{R}$.}
\label{fig:RDI}
\end{figure}

The resulting $\Omega h^{2}$ as a function of the DM mass, for different values of  $v_{R}= 2, 5, 10$~TeV, is shown in Fig.~\ref{fig:RDI}. The current bound provided by Planck \cite{Ade:2013zuv}:
\begin{equation}
\Omega h^{2} = 0.1199 \pm 0.0027 \, ,
\label{plankbound}
\end{equation} 
gives important restrictions on the parameter space of our model. As expected, there is a strong dependence of the relic density on $M_{W_{R}}$ and $M_{Z_{R}}$ which appear due to the (co-) annihilation channels involving $W_{R}$ and $Z_{R}$. The dips in the figure around $M_{\text{DM}}\simeq M_{W_{R}}/2$ and $M_{\text{DM}}\simeq M_{Z_R}/2$ correspond to the $W_{R}$ and the $Z_{R}$ resonances respectively. As one can see, for each value of $v_{R}$, the coannihilation effects are most important for the region where the DM mass is below the first resonance, i.e. $M_{\text{DM}}\leq M_{W_R}/2$. On the other hand, for values of $M_{\text{DM}}$ above the second resonance, the annihilation effects become less important and the relic density increases. The most important contributions to the relic density come from the channels $\Psi \overline{\Psi} \rightarrow W_{R} \gamma$, $\Psi \overline{\Psi} \rightarrow q \overline{q}$ and $\Psi \overline{\Psi} \rightarrow W^{+}_{R}W^{-}_{R}$ via the exchange of $\Psi$, $W_{R}$ and $Z_{R}$ respectively. Note that for $M_{\text{DM}} \gsim 2$ TeV, the correct relic density can be obtained only if $M_{\text{DM}} \lsim M_{W_R},M_{Z_R}$.


\begin{figure}[htb]
\includegraphics[width=0.6\textwidth]{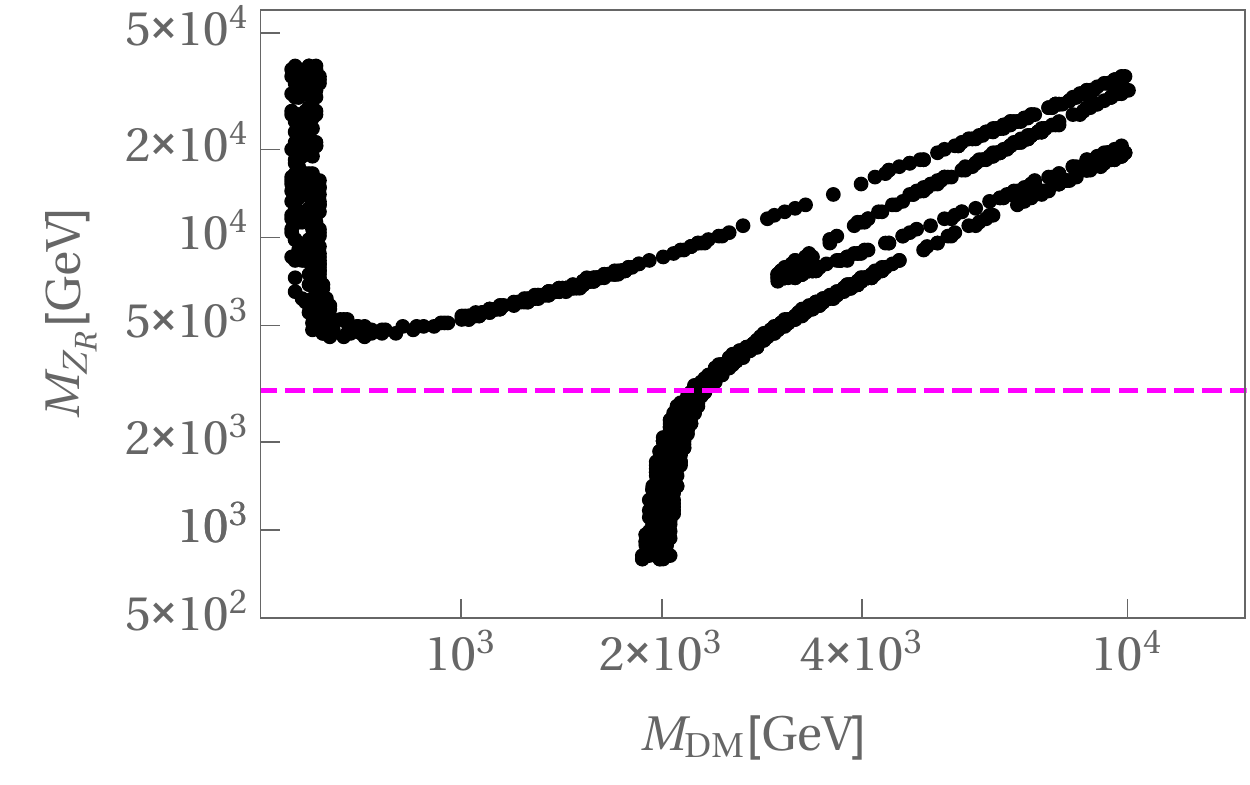}  %
\caption{Allowed values of $M_{\text{DM}}$ vs  $M_{Z_R}$, when the DM relic density is in the $3\sigma$-range of the relic density in Eq.~\eqref{plankbound}. The dashed line shows the lower limit on $M_{Z_{R}}$ imposed by CMS \cite{Khachatryan:2016zqb}}.
\label{fig:PSI}
\end{figure}

Now, if instead of fixing $v_{R}$, we let this scale as a free parameter in the range of $0.5< v_{R}/\text{TeV}<50$, the allowed region imposed by Planck in the plane spanned by the DM mass $M_{\text{DM}}$ and the $Z_{R}$ mass $M_{Z_R}$ is shown in Fig.~\ref{fig:PSI}. As we can appreciate, there is a region of points which is associated with the $Z_{R}$ and $W_{R}$ resonances for $M_{Z_R}\gtrsim 7~\text{TeV}$. For the lowest values of $M_{\text{DM}}$, only large values of $M_{Z_R}$ are allowed, for example, for $M_{\text{DM}}\simeq 700$~GeV, $M_{Z_R}\simeq [7, 40]$~TeV. As observed also from Fig.~\ref{fig:RDI}, for larger values of $M_{Z_R}$, larger values of $M_{\text{DM}}$ are allowed. Importantly, values of $M_{\text{DM}}\gtrsim 10$ TeV  are ruled out by the bound given in Eq.~\eqref{plankbound}. In addition, the current LHC limit of approximately $M_{Z_{R}}\gtrsim 3$ TeV \cite{Khachatryan:2016zqb}, based on the first few fb$^{-1}$ of the 2016 data set, excludes part of the otherwise allowed range of $M_{\text{DM}}$ in the region of $M_{\text{DM}}\simeq 2$ TeV. We expect that the updated analysis of the full $\sqrt{s}=13$ TeV data set will increase this limit towards  $M_{Z_{R}}\gsim 5$ TeV. 

As mentioned before, due to the fact that in this simple scenario our DM is a Majorana particle with hypercharge zero, there is no direct $Z$-exchange and we have a  zero tree-level $\sigma_{N}^{\text{SI}}$. Due to this fact, this scenario is nearly entirely unconstrained by direct detection experiments. However, as will be described in the next section, adding an extra fermionic field to the DM setup, opens a LR scalar portal  and then a non-vanishing $\sigma_{N}^{\text{SI}}$ arises.

\subsection{Case II}

\begin{figure}
\def\fh{5cm}
\includegraphics[height=\fh]{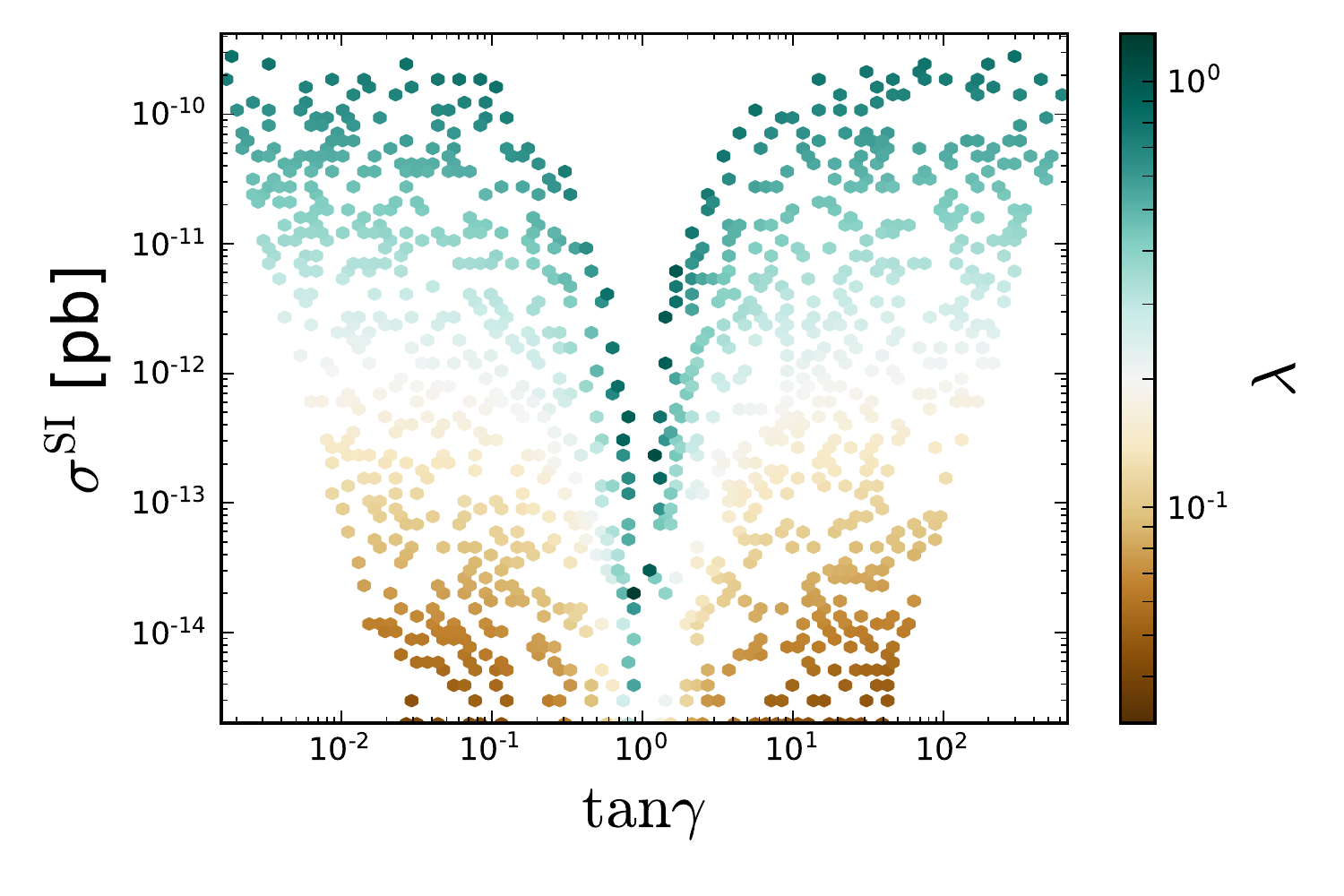}
\includegraphics[height=\fh,width=6cm]{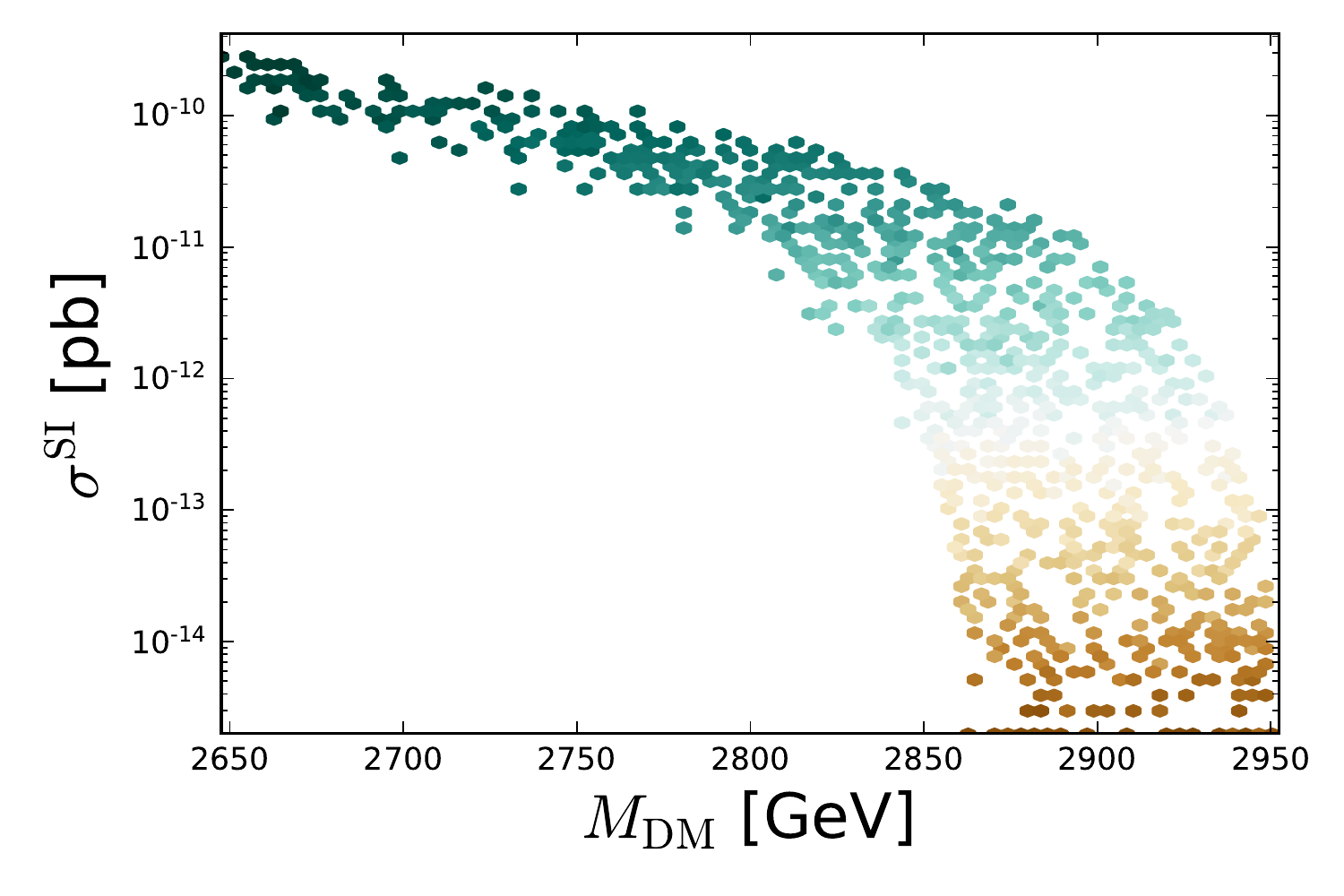}
\caption{Direct detection cross section for  $v_R=6\ \text{TeV}$ and $M_{11}=50\ \text{TeV}$. The color variation of $\lambda$ is the same for both plots.}
\label{fig:lam}
\end{figure}

In this scenario we introduce an extra Weyl fermion, $\Psi_{1130}$ in addition  to the DM setup described in \emph{Case~I}. As in the previous scenario, due the absence of any interaction terms mediating the decay of $\Psi_{1130}$ into the SM particles, the lightest component of this triplet is accidentally stable and hence can be  a DM candidate. A mixture of the neutral components of the fields $\Psi_{1130}\oplus \Psi_{1132} \oplus {\bar\Psi}_{113-2}$ is the DM. The relevant mass terms and interactions of the new fields, including the scalar portal, are given by:
\begin{align}
L\supset & M_{11}\operatorname{Tr}(\Psi_{1130}\Psi_{1130})+M_{23}\operatorname{Tr}(\Psi_{1132}{\bar\Psi}_{113-2})\nonumber\\
&+\lambda_{13}\operatorname{Tr}(\Delta_{R}{\bar\Psi}_{113-2}\Psi_{1130})+\lambda_{12}\operatorname{Tr}(\Delta^{\dagger}_{R}\Psi_{1132} \Psi_{1130})\, ,
\end{align}
where $M_{11}$ is the tree level mass of the components of  the triplet $\Psi_{1130}$. The LR scalar ``portal" is given by the interactions of the new field $\Psi_{1130}$ with the scalar boson $\Delta_{R}$. These interactions are proportional to the $\lambda_{13}$ and $\lambda_{12}$  Yukawa couplings. Depending on the choice of these Yukawa parameters, the direct detection nucleon cross section, $\sigma_{N}^{\text{SI}}$, will or will not be different from zero. We define
\begin{align}
\tan\gamma=&\frac{\lambda_{13}}{\lambda_{12}}\,,& \lambda=&\sqrt{\lambda_{12}^2+\lambda_{13}^2}\,.
\end{align}
To illustrate the dependence of the direct detection cross section, $\sigma_{N}^{\text{SI}}$, on these parameters, we choose a point with a $v_R=6\ \text{TeV}$ and $M_{11}=50\ \text{TeV}$. We then scan over the other parameters as illustrated in Fig~\ref{fig:lam}, with $2.7<M_{23}/\text{TeV}<3.1$. We can see that $\sigma_{N}^{\text{SI}}$ is proportional to $\lambda$. As illustrated in the left panel of Fig.~\ref{fig:lam}, there is a blind spot\footnote{The blind spot corresponds to the zone in the parameter space where the coupling between the DM and the scalar sector is zero, leading to vanishing direct detection cross section.} for positive values of $\tan\gamma$ at $\tan\gamma=1$. This is expected, since for decoupled $M_{11}$, the mixing with the scalar is proportional to $M_{23}\sin2\gamma - M_{\text{DM}}$, with  $M_{\text{DM}}\approx M_{23}$. Note that values for $|\tan\gamma|>1$ are equivalent to the values  with $|\tan\gamma|<1$.  In what follows we only consider the region $|\tan\gamma|\ge 1$.  In the right  panel of Fig.~\ref{fig:lam},  we show explicitly the dependence of  $\sigma_{N}^{\text{SI}}$ with $M_{\text{DM}}$ for the same color-range of $\lambda$. There, we include only  points well outside the blind spot with $\tan\gamma>5$.
Note that it is sufficient that only one of the Yukawa couplings $\lambda_{12}$ or $\lambda_{13}$ is different from zero to obtain a non-vanishing $\sigma_{N}^{SI}$.\\ 
In our scans, we choose the Yukawa couplings $\lambda_{12}$ and
$\lambda_{13}$ to be small, in the ballpark $(\lambda_{12} ,
\lambda_{13} \leq 0.1)$. Since the RGEs for $\lambda$ are proportional 
to the $\lambda$'s themselves, we expect that also a the GUT scale
these couplings remain perturbative, i.e: $(\lambda_{12} (mG ),
\lambda_{13} (mG )) \leq 4\pi$.

\begin{figure}
\includegraphics[width=0.65\textwidth]{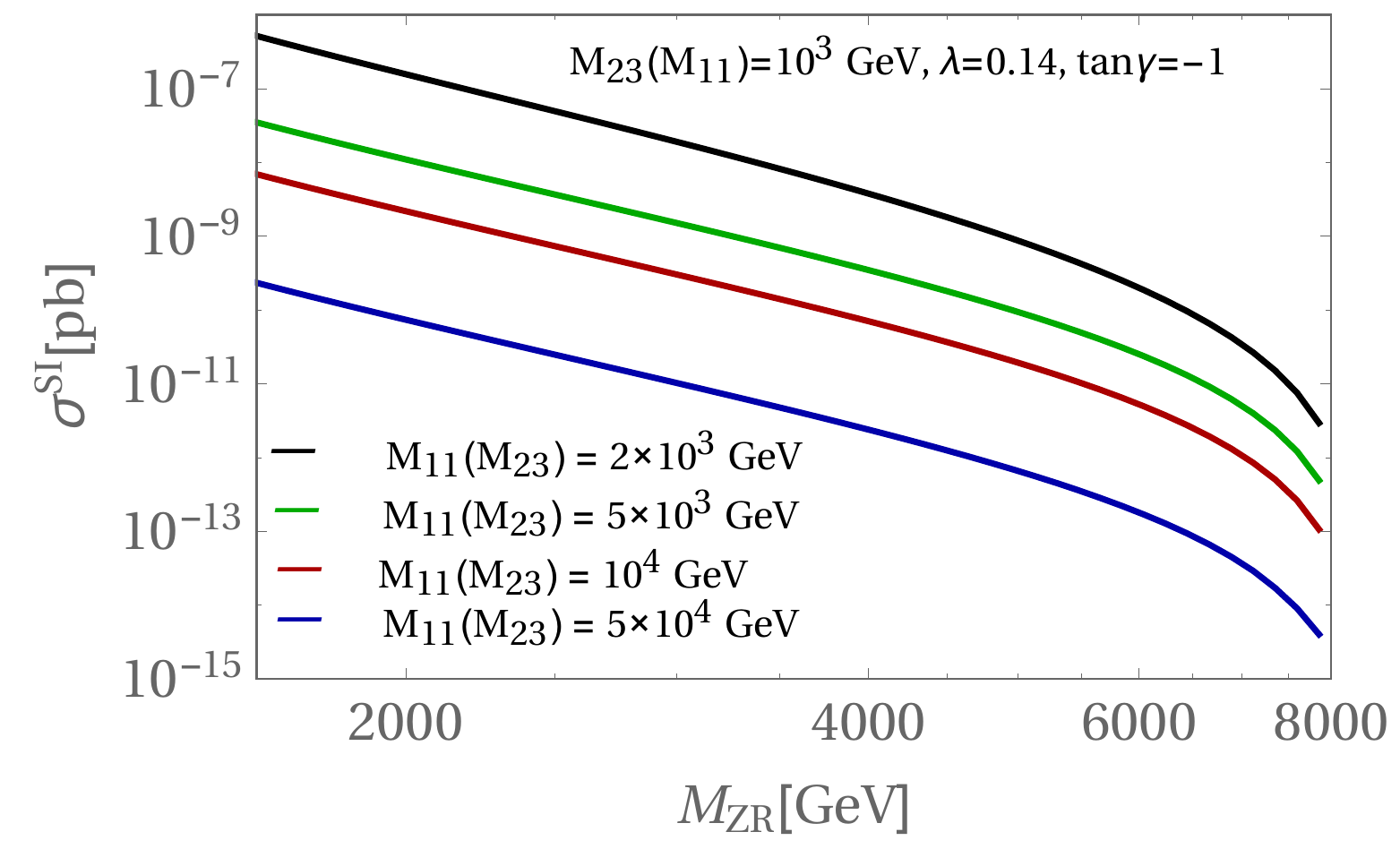} 
\caption{Direct detection rate vs $M_{Z_R}$ for $M_{23}, M_{11}=10^3$~GeV and different values of $M_{11}$ and $M_{23}$ respectively}
\label{fig:DD1}
\end{figure}

In Fig.~\ref{fig:DD1} $\sigma^{\text{SI}}_{N}$ is shown as a function of  $M_{Z_R}$ for different values of $M_{\text{DM}}$,  without imposing the constraint from the proper relic density. The curves correspond to different choices of $M_{11}$ ($M_{23}$)  for fixed values of $M_{23}=1$~TeV ($M_{11}=1$~TeV), $\tan\gamma=-1$, and $\lambda=0.14$.  As expected, $\sigma^{\text{SI}}_N$ decrease as $M_{11}$  ($M_{23}$) increase, recovering back the simplest \emph{Case~I} when $M_{11}$ is sufficiently high.

\begin{figure}[htb]
\hspace{-0.5cm}
\includegraphics[width=0.5\textwidth]{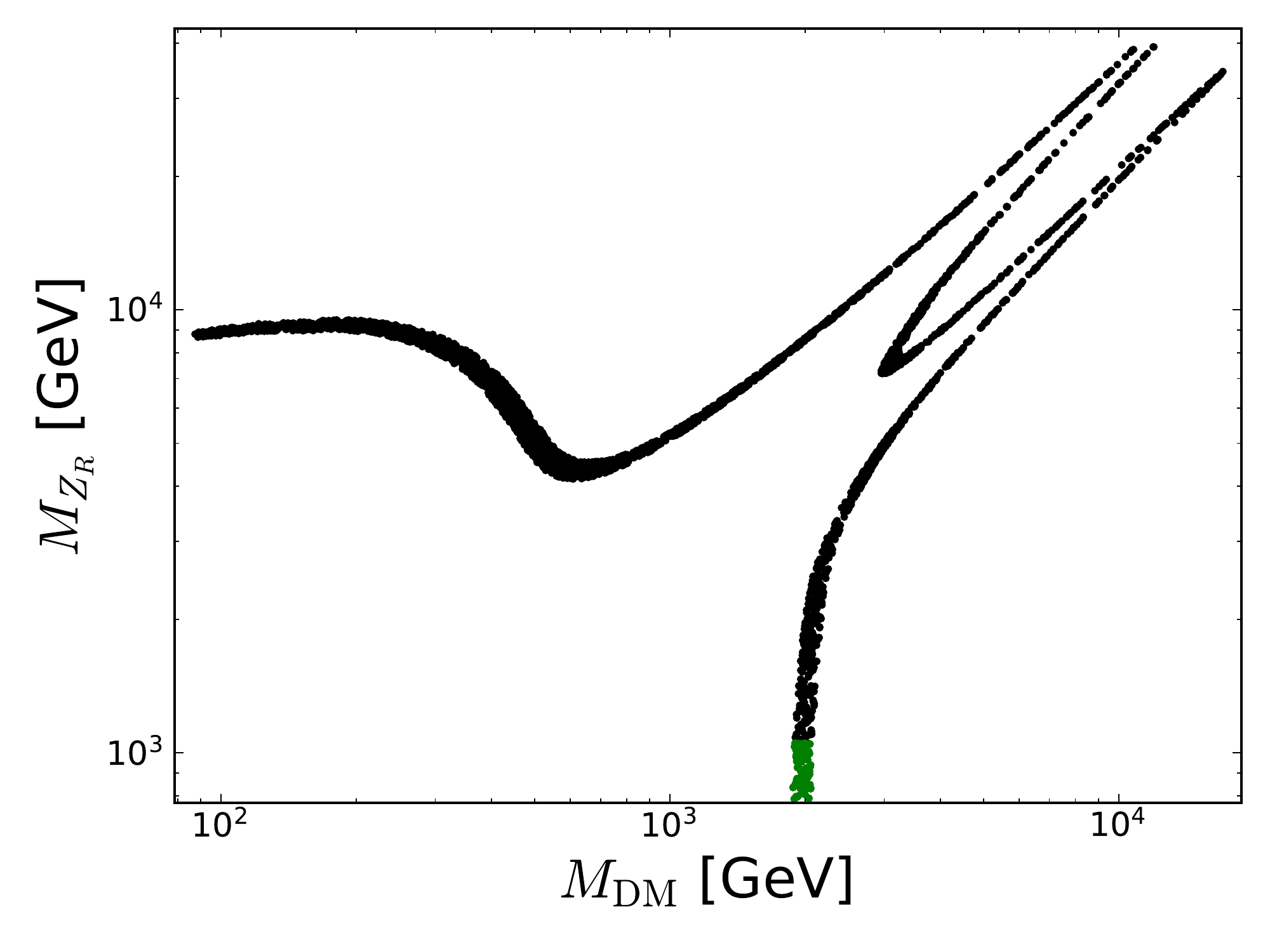}
\begin{tabular}{cc}
\includegraphics[width=0.5\textwidth]{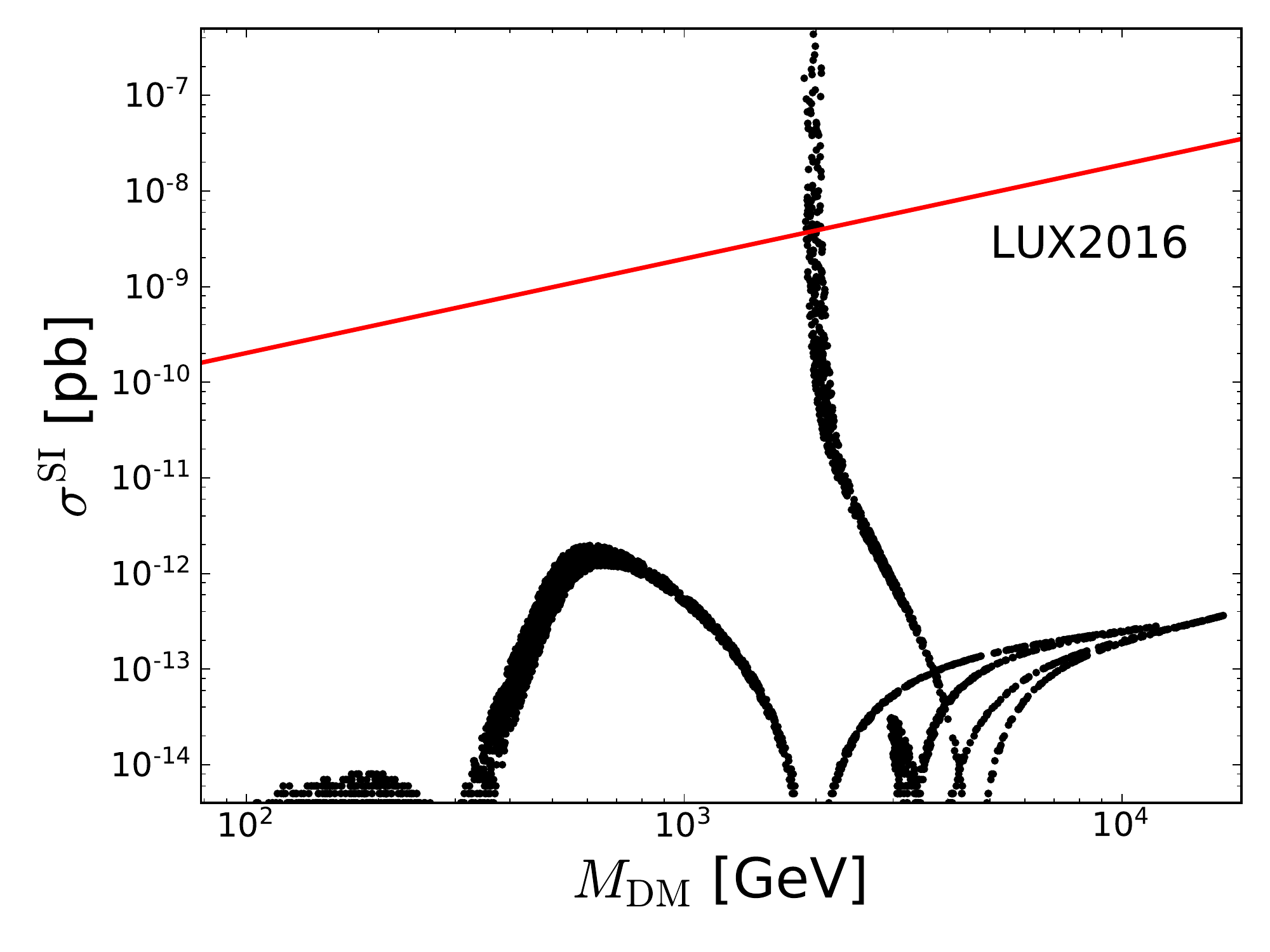} & %
\includegraphics[width=0.5\textwidth]{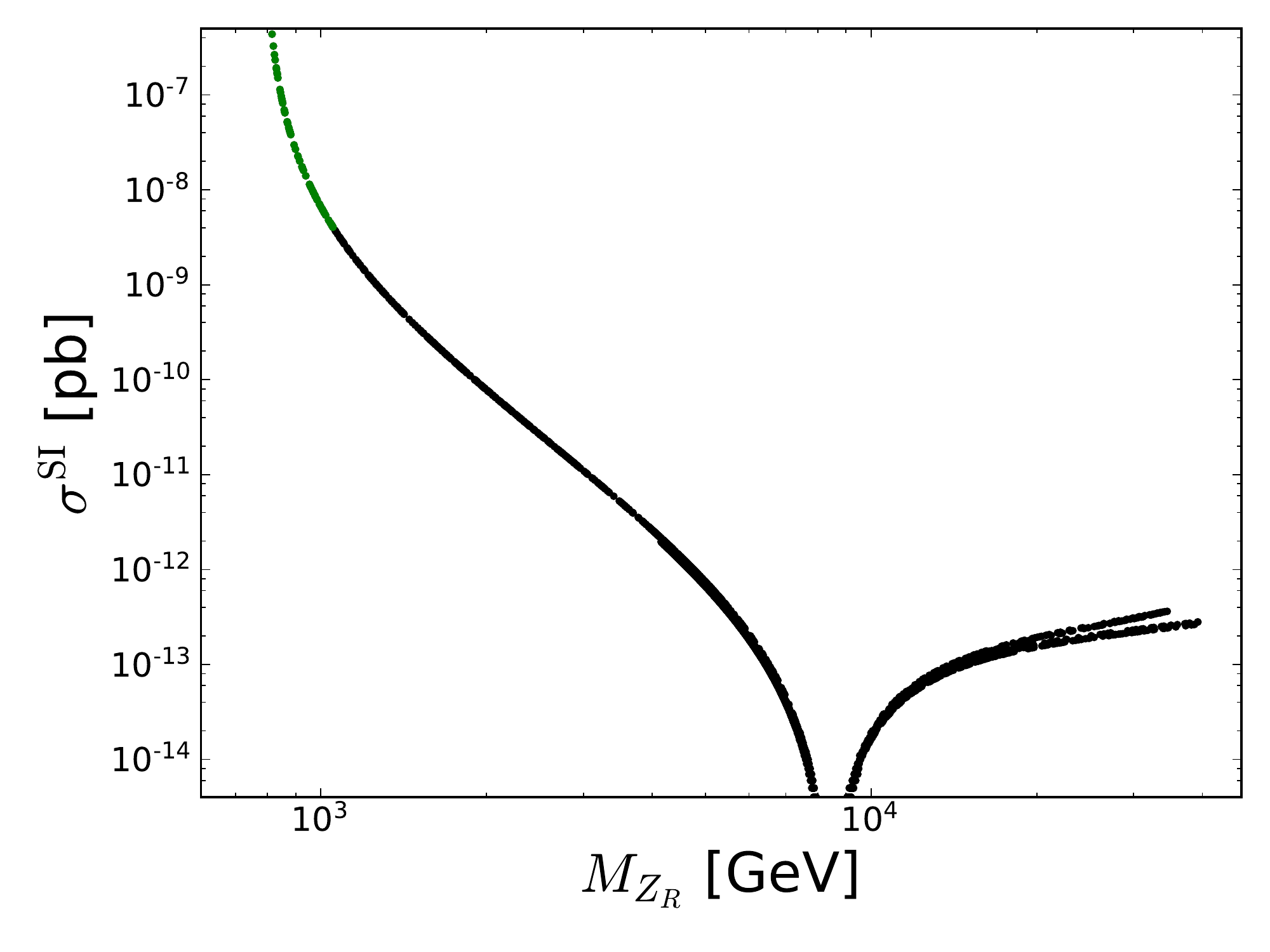}%
\end{tabular}
\caption{Allowed parameter space for  $v_{R}: [2, 50]$~TeV, $M_{23}:[0.2, 50]$~TeV, $M_{11}: 50 $~TeV, $\tan\gamma=-1$, and   $\lambda=0.14$. The green points in the upper panel correspond to the $M_{\text{DM}}$ and $M_{Z_{R}}$ masses excluded by the LUX bounds.}
\label{fig:PS2}
\end{figure}

The allowed parameter space restricted by the relic density Planck bound Eq.~\eqref{plankbound}, in the plane spanned by $M_{\text{DM}}$ and  $M_{Z_R}$, is shown in the upper panel of Fig.~\ref{fig:PS2} for a specific choice of the  parameters: $v_{R}: [2, 50]$ TeV, $M_{23}:[0.2, 50]$ TeV, $M_{11}: 50$ TeV,  $\tan\gamma=-1$ and $\lambda=0.14$. The scan includes the case  $M_{23}\ll M_{11}$, approaching then the simple DM scenario described in \emph{Case I}, where $\sigma^{\text{SI}}_{N}$ is zero.  Hence the similarity between both plots. Note however that the mixing opens up the window of small DM masses when $M_{\text{DM}}<M_{23}$.  
Moreover, the region of low $M_{Z_R}$ corresponding to the green points in the plot,  are excluded by the spin-independent elastic DM-nucleon direct detection constraints   from  LUX-2016  bound~\cite{Akerib:2013tjd}.

The numerical results for this constraints are shown explicitly in the lower panel of Fig.~\ref{fig:PS2} as a function of  $M_{\text{DM}}$ (left) and  $M_{Z_R}$ (right).  From the left down panel, we can appreciate that the LUX-2016  bound  on $\sigma^{\text{SI}}_{N}$, is above almost all the points in the plane spanned by $\sigma^{\text{SI}}$ and $M_{\text{DM}}$  allowing DM into the range $M_{\text{DM}}\sim[0.1, 10]$~TeV, except for one small window around $M_{\text{DM}}\sim 2$~TeV.  From the down right panel we can also observe that the LUX-2016 bound significantly cuts the parameter space for low $M_{Z_R}$, and allows only $Z_{R}$ masses larger than about $M_{Z_R}\sim 1$~TeV. Future limits from direct detection might lead to constraints that are competitive to the colliders limits for this scenario. It is expected that the projected values for XENON \cite{Aprile:2012nq,Aprile:2012zx} impose more stringent constraints in the values of $M_{\text{DM}}$ and $M_{Z_R}$. It is worth noting that the $M_{Z_R}$ current limit given by the LHC \cite{Khachatryan:2014dka} $M_{Z_R}\geq [2.6-3.5]$~TeV, depending on the $Z_R$ couplings (i.e. the values of $g_R$ and $g_{B-L}$, makes this scenario consistent with the relic density constraints and the XENON100 and the LUX bounds.

%
%
The allowed values of the DM restricted by direct detection analysis are in perfectly agreement with range of DM masses which fulfill all the GUT phenomenological requirements, presented in Sec.~\ref{sec:gut}.

\section{Conclusions}
\label{sec:conclusions}

We explored simple left right scenarios with a dark matter candidate as a mixed state of fermionic $SU(2)_{R}$ triplets. Such models, denoted as \emph{Case~I} and \emph{Case~II}, correspond to combinations of triplet-triplet and triplet-triplet-triplet DM candidates respectively, not explored in the literature for a wide range of DM masses. Acceptable relic abundance, imposed by the Planck bound, is obtained for a wide range of masses in each of the models. Due to the  Majorana nature of the DM and the absence of $VV$ interactions, a vanishing tree-level cross section $\sigma^{\text{SI}}_{N}$ is obtained in \emph{Case I}. This model is less constrained than models with left right triplet-triplet DM candidates~\cite{Garcia-Cely:2015quu,Heeck:2015qra}. 
A non zero $\sigma^{\text{SI}}_{N}\ne0$  appears in \emph{Case~II} through the  interactions with the LR scalar sector. 
The direct detection parameter space in \emph{Case II} is constrained by the bounds imposed by the LUX-2016 results in a competitive way with collider constraints. More stringent constraints are expected from future experiments such as XENON1T.
The unification of the gauge couplings through the $SO(10)$-LR channel $SU(3)\times SU(2)_{L}\times SU(2)_{R}\times U(2)_{B-L}$ is achieved in our model, by requiring some additional fermionic fields up to the SM and DM setup. Part of the parameter space allowed by the DM bounds is perfectly compatible with the parameter space  which fulfill all the GUT phenomenological constraints.

\section*{Acknowledgments}

C.A. acknowledges support by CONICYT  (Chile)  Ring  ACT1406 and Basal FB0821. D.R. has been partially supported by UdeA through the grants Sostenibilidad-GFIF,  and COLCIENCIAS through the Grant No. 111-565-842691. M.H. is supported by  Spanish  MICINN grants SEV-2014-0398, FPA2014-58183-P, Multidark CSD2009-00064 (MINECO) and PROMETEOII/2014/084 (Generalitat Valenciana).
The autors thank Werner Porod and Florian Staub for helpful assistance with the SARAH/SPHENO and toolbox packages.

\appendix
\section{}

Under $SO(10)$, a whole family of SM quarks and fermions belongs to the {\bf 16} representation which is $3(B-L)$ odd. On the other hand, the SM Higgs, which belong to the  {\bf 10} $SO(10)$ representation is even. As a result, if all the fields breaking $U(1)_{B-L}$ and $SO(10)$ are $3(B-L)$ even, a $Z_{2}$ symmetry will remain unbroken. This lead two possible stable DM candidates: scalar DM which has to belong to a $SO(10)$ representation odd under $3(B-L)$, because all the other scalar particle combinations it couple to, or decay to, are even. On the other hand, DM could be a fermion if it belongs to $SO(10)$ representation even under $3(B-L)$, because all the other fermion combination it couples to or decay to are odd \cite{Hambye:2010zb}. Considering also that DM must be colorless and electrically neutral, the different possibilities of scalar and fermionic DM candidates under $SO(10)$ are depicted in Table~\ref{tab:Reps}.

\begin{table}[htb]
\begin{tabular}{|l|l|l|l|l|l|l|l|l|l|l|l|l|l|l|l|l|l|l|l|}
\hline
\hline 
$SO(10)$ & \multicolumn{2}{ c| }{${\bf 16}$} & \multicolumn{4}{ c| }{${\bf 144}$}&  \multicolumn{2}{ c| }{${\bf 10}$} & \multicolumn{3}{ c| }{\bf 45} & \multicolumn{2}{ c| }{\bf 54} & \multicolumn{1}{ c| }{${\bf 120}$} & \multicolumn{3}{ c| }{${\bf 126}$} & \multicolumn{1}{ c| }{${\bf  \overline{126}}$} \\ [2ex]
\hline          
$SU(3)_{c}$ & 1 & 1 & 1 & 1 & 1 & 1 & 1 &  1 & 1 & 1 & 1 & 1 &1 & 1  &1 &1 &1 &1  \\ [3ex]
$SU(2)_{L}$ & 1 & 2 & 2 & 1 & 3 & 2 & 2& 1& 1& 1& 3 & 1 & 3 &2 & 2& 3&1 & 1   \\ [3ex]
$SU(2)_{R}$ & 2 & 1 & 1 & 2 & 2 & 3& 2&1 & 3& 1 & 1 & 1 &3 &2 &2 &1 & 3 &3  \\ [3ex]
$U(1)_{B-L}$ & -1 & 1 & 1 & -1 & -1 & 1  & 0&0 &0 &0 & 0 & 0 &0 & 0 &0 & -2 &2 & -2  \\ [3ex]
\hline
$U(1)_{Y}$ & 0 & $-\frac{1}{2}$ & $-\frac{1}{2}$ & 0 & 0 & $-\frac{1}{2}$& $-\frac{1}{2}$ & 0 & 0 & 0 & 0 & 0 &-1 & $-\frac{1}{2}$ &$-\frac{1}{2}$ & -1 &0 & 0 \\ [3ex]
\hline
Scalar DM  & \checked &  \checked  & \checked & \checked & \checked & \checked & &  &  & & & &  & & & & &   \\ [2ex] 
Fermion DM & & & & & & & \checked&  \checked & \checked & \checked & \checked &\checked &\checked & \checked & \checked & \checked & \checked& \checked  \\ [2ex] 
\hline
\hline
\end{tabular}
\caption{Different dark matter candidates coming from  $SO(10)$ representations up to ${\bf 126}$.}
\label{tab:Reps}
\end{table}

\section{}

The equation for the running of the inverse gauge couplings at 1-loop lavel can be written as:

\begin{equation}
\alpha^{-1}_{i}(t)=\alpha_{i}^{-1}(t_{0})+\frac{b_{i}}{2\pi}(t-t_{0})
\end{equation}

where $t_{i}=\text{log}(m_{i})$, as usual. The effective one-loop $\beta-$RGE coefficients are given by:

\begin{align}
(b^{SM}_{3}, b^{SM}_{2}, b^{SM}_{1})&=(-7, -19/6, 41/10) \noindent \\
(b^{SM}_{3}, b^{SM}_{2}, b^{SM}_{1})&=(-7, -3, -3, 4)+(\Delta b_{3}^{LR}, \Delta b_{2}^{LR}, \Delta b^{LR}_{R}, \Delta b^{LR}_{B-L})
\end{align}

and the (B-L) charges are written in the canonical normalization. The contributions from the additional scalar and fermionic fields in the regime: $[m_{LR}, m_{G}]$, not accounted for in the SM are given by:

\begin{equation}
(\Delta b_{3}^{LR}, \Delta b_{2}^{LR}, \Delta b_{R}^{LR}, \Delta b_{(B-L)}^{LR}) =(10/3, 10/3, 14/3, 47/6) 
\end{equation}

\bibliographystyle{h-physrev4}
\bibliography{darkmatter}

\begin{thebibliography}{10}

\bibitem{Belanger:2012zr}
G.~Belanger, K.~Kannike, A.~Pukhov, and M.~Raidal,
\newblock JCAP {\bf 1301}, 022 (2013), arXiv:1211.1014.

\bibitem{Belanger:2012vp}
G.~Belanger, K.~Kannike, A.~Pukhov, and M.~Raidal,
\newblock JCAP {\bf 1204}, 010 (2012), arXiv:1202.2962.

\bibitem{Hirsch:2010ru}
M.~Hirsch, S.~Morisi, E.~Peinado, and J.~W.~F. Valle,
\newblock Phys. Rev. {\bf D82}, 116003 (2010), arXiv:1007.0871.

\bibitem{Boucenna:2011tj}
M.~S. Boucenna {\em et~al.},
\newblock JHEP {\bf 05}, 037 (2011), arXiv:1101.2874.

\bibitem{Kadastik:2009dj}
M.~Kadastik, K.~Kannike, and M.~Raidal,
\newblock Phys. Rev. {\bf D81}, 015002 (2010), arXiv:0903.2475.

\bibitem{Kadastik:2009cu}
M.~Kadastik, K.~Kannike, and M.~Raidal,
\newblock Phys. Rev. {\bf D80}, 085020 (2009), arXiv:0907.1894,
\newblock [Erratum: Phys. Rev.D81,029903(2010)].

\bibitem{Frigerio:2009wf}
M.~Frigerio and T.~Hambye,
\newblock Phys. Rev. {\bf D81}, 075002 (2010), arXiv:0912.1545.

\bibitem{Senjanovic:1975rk}
G.~Senjanovic and R.~N. Mohapatra,
\newblock Phys. Rev. {\bf D12}, 1502 (1975).

\bibitem{Brahmachari:1991np}
B.~Brahmachari, U.~Sarkar, and K.~Sridhar,
\newblock Phys. Lett. {\bf B297}, 105 (1992).

\bibitem{Arbelaez:2013nga}
C.~Arbeláez, M.~Hirsch, M.~Malinský, and J.~C. Romão,
\newblock Phys. Rev. {\bf D89}, 035002 (2014), arXiv:1311.3228.

\bibitem{Martin:1992mq}
S.~P. Martin,
\newblock Phys. Rev. {\bf D46}, R2769 (1992), arXiv:hep-ph/9207218.

\bibitem{Mambrini:2016dca}
Y.~Mambrini, N.~Nagata, K.~A. Olive, and J.~Zheng,
\newblock Phys. Rev. {\bf D93}, 111703 (2016), arXiv:1602.05583.

\bibitem{Heeck:2015qra}
J.~Heeck and S.~Patra,
\newblock Phys. Rev. Lett. {\bf 115}, 121804 (2015), arXiv:1507.01584.

\bibitem{Agarwalla:2016rmw}
S.~K. Agarwalla, K.~Ghosh, and A.~Patra,
\newblock (2016), arXiv:1607.03878.

\bibitem{Garcia-Cely:2015quu}
C.~Garcia-Cely and J.~Heeck,
\newblock (2015), arXiv:1512.03332,
\newblock [JCAP1603,021(2016)].

\bibitem{Boucenna:2015sdg}
S.~M. Boucenna, M.~B. Krauss, and E.~Nardi,
\newblock Phys. Lett. {\bf B755}, 168 (2016), arXiv:1511.02524.

\bibitem{Arbelaez:2015ila}
C.~Arbelaez, R.~Longas, D.~Restrepo, and O.~Zapata,
\newblock Phys. Rev. {\bf D93}, 013012 (2016), arXiv:1509.06313.

\bibitem{Nagata:2016knk}
N.~Nagata, K.~A. Olive, and J.~Zheng,
\newblock (2016), arXiv:1611.04693.

\bibitem{Dey:2015bur}
U.~K. Dey, S.~Mohanty, and G.~Tomar,
\newblock Phys. Lett. {\bf B756}, 384 (2016), arXiv:1512.07212.

\bibitem{Mambrini:2015vna}
Y.~Mambrini, N.~Nagata, K.~A. Olive, J.~Quevillon, and J.~Zheng,
\newblock Phys. Rev. {\bf D91}, 095010 (2015), arXiv:1502.06929.

\bibitem{Dev:2016xcp}
P.~S. Bhupal~Dev, R.~N. Mohapatra, and Y.~Zhang,
\newblock JHEP {\bf 11}, 077 (2016), arXiv:1608.06266.

\bibitem{Dev:2016qeb}
P.~S.~B. Dev, R.~N. Mohapatra, and Y.~Zhang,
\newblock (2016), arXiv:1610.05738.

\bibitem{Berlin:2016eem}
A.~Berlin, P.~J. Fox, D.~Hooper, and G.~Mohlabeng,
\newblock JCAP {\bf 1606}, 016 (2016), arXiv:1604.06100.

\bibitem{Patra:2015qny}
S.~Patra,
\newblock Phys. Rev. {\bf D93}, 093001 (2016), arXiv:1512.04739.

\bibitem{Berlin:2016hqw}
A.~Berlin,
\newblock Phys. Rev. {\bf D93}, 055015 (2016), arXiv:1601.01381.

\bibitem{PhysRevD.22.2227}
J.~Schechter and J.~W.~F. Valle,
\newblock Phys. Rev. D {\bf 22}, 2227 (1980).

\bibitem{Mohapatra:1986uf}
R.~N. Mohapatra,
\newblock {\em {UNIFICATION AND SUPERSYMMETRY. THE FRONTIERS OF QUARK - LEPTON
  PHYSICS}} (Springer, Berlin, 1986).

\bibitem{DeRomeri:2011ie}
V.~De~Romeri, M.~Hirsch, and M.~Malinsky,
\newblock Phys. Rev. {\bf D84}, 053012 (2011), arXiv:1107.3412.

\bibitem{Arbelaez:2013hr}
C.~Arbelaez, R.~M. Fonseca, M.~Hirsch, and J.~C. Romao,
\newblock Phys. Rev. {\bf D87}, 075010 (2013), arXiv:1301.6085.

\bibitem{Nath:2006ut}
P.~Nath and P.~Fileviez~Perez,
\newblock Phys. Rept. {\bf 441}, 191 (2007), arXiv:hep-ph/0601023.

\bibitem{Abe:2013lua}
Super-Kamiokande, K.~Abe {\em et~al.},
\newblock Phys. Rev. Lett. {\bf 113}, 121802 (2014), arXiv:1305.4391.

\bibitem{Staub:2010jh}
F.~Staub,
\newblock Comput. Phys. Commun. {\bf 182}, 808 (2011), arXiv:1002.0840.

\bibitem{Staub:2008uz}
F.~Staub,
\newblock (2008), arXiv:0806.0538.

\bibitem{Staub:2009bi}
F.~Staub,
\newblock Comput. Phys. Commun. {\bf 181}, 1077 (2010), arXiv:0909.2863.

\bibitem{Staub:2016dxq}
F.~Staub {\em et~al.},
\newblock Eur. Phys. J. {\bf C76}, 516 (2016), arXiv:1602.05581.

\bibitem{Porod:2003um}
W.~Porod,
\newblock Comput. Phys. Commun. {\bf 153}, 275 (2003), arXiv:hep-ph/0301101.

\bibitem{Porod:2011nf}
W.~Porod and F.~Staub,
\newblock Comput. Phys. Commun. {\bf 183}, 2458 (2012), arXiv:1104.1573.

\bibitem{Belanger:2006is}
G.~Belanger, F.~Boudjema, A.~Pukhov, and A.~Semenov,
\newblock Comput. Phys. Commun. {\bf 176}, 367 (2007), arXiv:hep-ph/0607059.

\bibitem{Belanger:2010st}
G.~Belanger, N.~D. Christensen, A.~Pukhov, and A.~Semenov,
\newblock Comput. Phys. Commun. {\bf 182}, 763 (2011), arXiv:1008.0181.

\bibitem{Staub:2011dp}
F.~Staub, T.~Ohl, W.~Porod, and C.~Speckner,
\newblock Comput. Phys. Commun. {\bf 183}, 2165 (2012), arXiv:1109.5147.

\bibitem{Ade:2013zuv}
Planck, P.~A.~R. Ade {\em et~al.},
\newblock Astron. Astrophys. {\bf 571}, A16 (2014), arXiv:1303.5076.

\bibitem{Khachatryan:2016zqb}
CMS, V.~Khachatryan {\em et~al.},
\newblock Phys. Lett. {\bf B768}, 57 (2017), arXiv:1609.05391.

\bibitem{Akerib:2013tjd}
LUX, D.~S. Akerib {\em et~al.},
\newblock Phys. Rev. Lett. {\bf 112}, 091303 (2014), arXiv:1310.8214.

\bibitem{Aprile:2012nq}
XENON100, E.~Aprile {\em et~al.},
\newblock Phys. Rev. Lett. {\bf 109}, 181301 (2012), arXiv:1207.5988.

\bibitem{Aprile:2012zx}
XENON1T, E.~Aprile,
\newblock Springer Proc. Phys. {\bf 148}, 93 (2013), arXiv:1206.6288.

\bibitem{Khachatryan:2014dka}
CMS, V.~Khachatryan {\em et~al.},
\newblock Eur. Phys. J. {\bf C74}, 3149 (2014), arXiv:1407.3683.

\bibitem{Hambye:2010zb}
T.~Hambye,
\newblock PoS {\bf IDM2010}, 098 (2011), arXiv:1012.4587.

\end{thebibliography}

\end{document}